\documentclass[5p,times]{elsarticle}
\usepackage[spanish,english]{babel}     %
\usepackage{lineno}
\usepackage{etoolbox}

\usepackage[T1]{fontenc}
\usepackage{caption}
\usepackage{textcomp}
\usepackage{amsmath}            
\usepackage{epstopdf}           
\usepackage{flushend}           
\usepackage{xspace}

\usepackage{amssymb}
\usepackage{amsthm}

\biboptions{comma,compress,super}

\usepackage[figuresright]{rotating}

\usepackage{subfigure}

\usepackage{float}
\spanishdecimal{.}
\begin{document}
\begin{frontmatter}
\title{Thermoelectric effects in self-similar multibarrier structure based on monolayer graphene}

\author[First]{M. Miniya}
\ead{m.miniya@uaem.edu.mx}
\author[Second]{O. Oubram}
\ead{oubram@uaem.mx}
\author[First]{A. G. Reynaud–Morales}
\ead{areynaud@uaem.mx}

\author[Third]{I. Rodríguez-Vargas \corref{cor2}}
\ead{isaac@uaz.edu.mx}

\author[First]{L. M. Gaggero-Sager \corref{cor1}}
\ead{lgaggero@uaem.mx}

\cortext[cor1]{lgaggero@uaem.mx}
\cortext[cor2]{isaac@uaz.edu.mx}
\address[First]{Centro de Investigación en Ingenierías y Ciencias Aplicadas, Universidad Autónoma Del Estado de Morelos, Av. Universidad 1001, Col. Chamilpa, 62209 Cuernavaca, Morelos, Mexico}
\address[Second]{Facultad de Ciencias Químicas e Ingeniería, Universidad Autónoma Del Estado de Morelos, Av. Universidad 1001, Col. Chamilpa, 62209 Cuernavaca, Morelos, Mexico}
\address[Third]{Unidad Acad\'emica de Ciencia y Tecnolog\'ia de la Luz y la Materia, Universidad Aut\'onoma de Zacatecas, Carretera Zacatecas-Guadalajara Km. 6, Ejido La Escondida, 98160 Zacatecas, Zac., Mexico}

\begin{abstract}
Thermoelectric effects have attracted wide attention in recent years from physicists and engineers. In this work, we explore the self-similar patterns in the thermoelectric effects of monolayer graphene based structures, by using the quantum relativistic Dirac equation. The transfer matrix method has been used to calculate the transmission coefficient. The Landauer–Büttiker formalism and the Cutler-Mott formula were used to calculate the conductance, the Seebeck coefficient, and the power factor. We find self-similar behavior and the scale factors between generations in the transport and thermoelectric properties. Furthermore, we implement these scale invariances as general scaling rules. We present a new analytical demonstration of self-similarity in the Seebeck coefficient. These findings can open outstanding perspectives for experimentalists to develop thermoelectric devices. 
\end{abstract}
\end{frontmatter}

\section{Introduction}
Graphene is a promising material in nanoelectronics \cite{geim2010rise}, due to its exotic properties and usefulness for understanding the physics of materials \cite{morozov2008electron}. Graphene is made out of carbon atoms arranged on a honeycomb structure \cite{neto2009electronic}. 
In the last decade, researchers have started to be interested in new structures to control the transport of electrons \cite{novoselov2004electric}. In specific, Dirac electrons can be controlled by the application of external electric and magnetic fields and/or substrates\cite{zhou2007substrate} in periodic or aperiodic fashion \cite{de2020magneto,dean2013hofstadter, sun2010transport}. These external effects modulate the band structure and consequently change the physical properties of the materials \cite{huard2007transport}. In principle, there are several structures, e.g., structures arranged following aperiodic sequences (Fibonacci, Thue-Morse, etc.) or self-similar patterns (Cantor set) . These systems have attracted great interest \cite{mukhopadhyay2010resonant, sena2010fractal, diaz2015scaling}, because of their exotic and unique electronic and optoelectronic properties \cite{lu2013transport}. Multibarrier systems in graphene, created by means of the different mechanisms indicated above are crucial to modulate the transport properties and probably the thermoelectric effects .  
\par Recently, fractal nanometric modulation has been reported \cite{gouyet1996physics}, according to the mathematical definition, fractals are homogeneous and self-similar geometric objects, which can be used to describe many physical phenomena \cite{cannon1984fractal}. Fractals exist everywhere \cite{barnsley2014fractals}, in certain plants such as romanesco broccoli, also on the coasts, the distribution of species, the growth of trees, rivers, and lungs. Many fractals have the property of self-similarity, at least approximately, if not exactly \cite{feder2013fractals}. A self-similar object is an object whose components resemble the whole shape. This reiteration of details or patterns occurs at progressively smaller scales, so that each part of a part, when enlarged, basically looks like a fixed part of the whole object. Indeed, a self-similar object remains invariant under scale changes, i.e., it has a symmetry of scale, so there is a scale factor which connects the similar parts. 
\par From an engineering perspective, fractal arrangements represent an important strategy for the integration of hard and soft materials  \cite{fan2014fractal, fairbanks2011fractal}, which allows adaptation to the increased elastic deformation of expandable electronics. Recently, a new fractal metamaterial based on graphene has been reported \cite{de2020graphene}. This new fractal metamaterial represents a propitious way for the realization of a broadband and compact platform for the future of optoelectronic devices. In addition, several studies \cite{diaz2016self, rodriguez2016self} have been carried out on the transmission of electrons in self-similar systems with graphene-based barriers. Another study reports a self-similar behavior in the conductance curves as well as mathematical expressions as scaling rules \cite{garcia2017self}. Transport properties with magnetic barriers have been reported in a graphene-based fractal system \cite{rodriguez2015role}. It is found that there are separations between energy minibands and regular oscillations in the conductance curves.  
\par In recent years, thermoelectric properties have attracted wide attention from researchers and engineers \cite{zevalkink2018practical}.  Thermoelectricity is a physical phenomenon which consists in the direct conversion of thermal energy into electricity \cite{hwang2009theory}. Latest research on thermoelectric effect has shown that it could be a solution to reduce the amount of heat that is wasted by today’s society. According to several studies, environmentally wasted heat represents more than $50 \%$ of the energy produced \cite{forman2016estimating}. Numerous theoretical studies have been carried out on the thermoelectric effect in various systems based on $2$D materials, showing an improvement in the thermoelectric performance \cite{dollfus2015thermoelectric,rodriguez2020self}.  
\par Many theoretical studies have been reported on structures based on low-dimensional materials, aimed to obtain higher values for the figure of merit ($ZT$) \cite{hicks1993effect, dresselhaus2007new}. However, in small dimensions quantum confinement occurs in this type of materials, which leads to redistribution of the density of states, which offers the possibility of varying power factor and thermal conductivity independently and ultimately improving the $ZT$ value. In other words, to increase the Seebeck coefficient, it is plausible that any effect that improve the energy dependence of the conductivity can be helpful. For example, by increasing the energy dependence of the charge density, which directly depends on the density of the states. In principle, $2$D systems provide much more conductivity which should be reflected on the Seebeck coefficient \cite{molina2020low}. 
\par In this paper, we have studied transport and thermoelectric effects in a new self-similar structure, based on graphene. In this system, the height of the barrier is scaled by the golden ratio. In fact, the golden number is among the few numbers that give the scaling rule. This kind of study has not been performed before, as it reports the self-similar patterns in the Seebeck coefficient and in the power factor. We consider the Dirac Hamiltonian to describe the behavior of the Fermi electrons through the graphene superlattice. The substrates such as SiO$_{2}$ and SiC, are introduced as a series of barriers-wells on the graphene monolayer. The transfer matrix formalism was used, as well as the Landauer-B\"{u}ttiker formalism, and the Cutler-Mott formula to calculate, for each generation, first the transmission coefficient, then the conductance, and finally the Seebeck coefficient and the power factor. We found self-similar patterns in electronic transport and in thermoelectric properties. Furthermore, we obtained the scaling rules that describe the invariance scale between generations of the system. We implemented these scaling rules as general mathematical expressions. We also carried out an analytical demonstration of scaling rules in the Seebeck coefficient, which has not been done either. Moreover, a comparative analysis between numerical and analytical scaling rules has been performed.   
\section{Methodology}
The self-similar thermoelectric system that we have considered in this work is formed by a graphene sheet, as shown schematically in Fig. 1. The system is constructed by rectangular potentials (barriers) and wells formed by substrates as SiC and SiO$_{2}$ arranged in a self-similar way along the propagation direction $x$-axis (see Fig. 2). 
In this system in particular, the height of the barrier is scaled by a golden ratio, represented by $\Phi$, with value  $\Phi=\frac{1+\sqrt{5}}{2}=1.61803398875$. In this respect, we studied multiple scaling factors in this kind of systems, however we find out that the golden ratio is one of the unique scaling factors that give scaling rules. Thus, the value of the scaling factor plays an important role, since it can affect drastically the scaling transformation among all generations. 
\\ The algorithm of self-similar barriers is created by taking a closed interval $[0, L_{x}]$, where $L_{x}$ is the total length of the system. We explain this algorithm in 2 steps:   
First step: Divide the closed interval in thirds [0, $L_{x}/3]\bigcup [L_{x}/3, 2L_{x}/3]\bigcup [2L_{x}/3, L_{x}]$ then put a barrier of length $L_{x}/3$ and energy $V_{0}$ in a middle interval, this is the main barrier of the system, the first generation ($N_{1}$) is obtained, (see Fig. 2).  
Second step: Divide the closed interval again in thirds $L_{x}/3^{2}$, repeat the previous process with the remaining intervals, putting barriers of length $L_{x}/3^{2}$ and energy $V_{0}/\Phi$. The main barrier is preserved and its width scaled by $L_{x}/3^{2}$, we get the second generation ($N_{2}$), (see Fig. 2). 
Repeating these steps with the barriers and segments of the second generation gives the third generation. It is important to note that to be self-similar barriers, the first generation should resemble each part of the subsequent generations. Therefore, every barrier in a given generation of the potential has a width given by $L_{x}/3^{N}$, where $N$ is the number of the generation. On the other hand, the energy of every set of barriers added in a given generation of the construction is scaled in its energy by a factor of $V_{0}/\Phi ^{N-1}$. These features of the potential are shown in Fig. 2, where it can be seen that this self-similar system is different from those analyzed previously \cite{diaz2015scaling,diaz2016self,rodriguez2016self}. 
\\From a theoretical standpoint, the influence of a substrate such as SiC (see Fig. 1), plays an important role in the physical properties of graphene, it enables to change the dispersion relation, from linear to parabolic, as well as to open a bandgap in the energy spectrum \cite{zhou2007substrate}. 
\\The structure can be studied with the Dirac equation:


\begin{figure}[!t] 
    \centering
    \includegraphics[width=7cm, height=4cm]{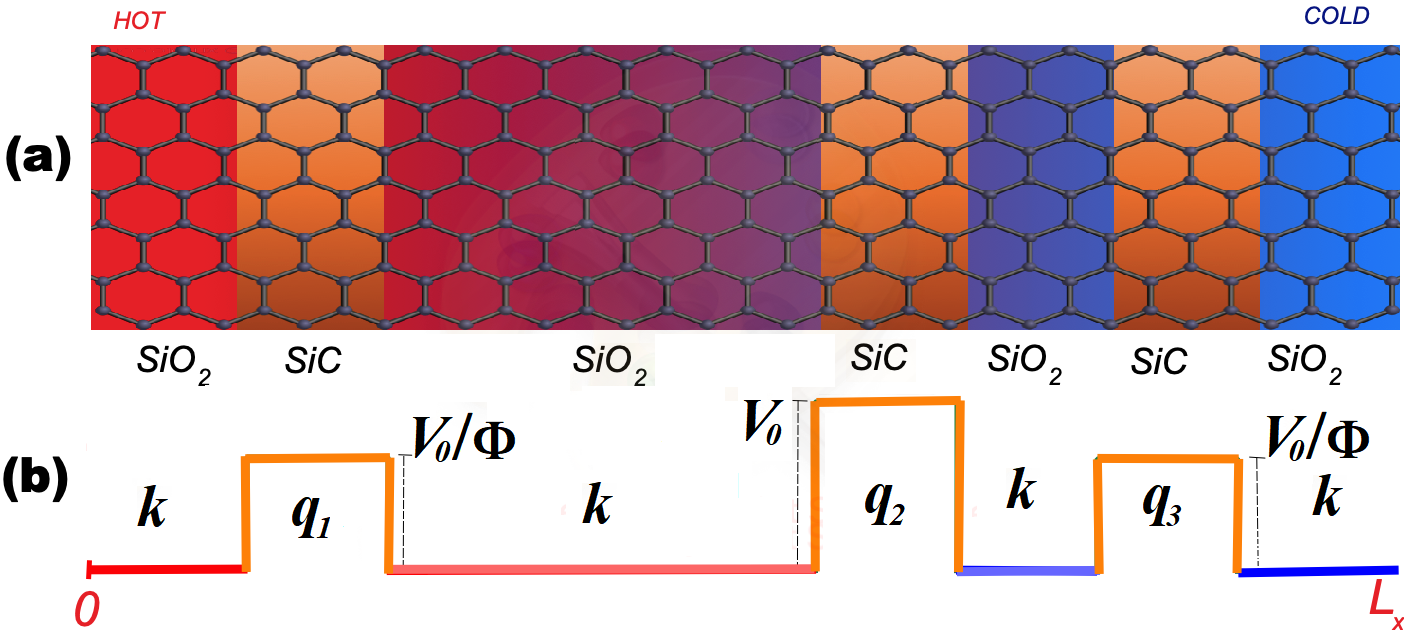}
    \caption{ \begin{normalsize} Graphical representation of the second generation $(N_{2})$ of a self-similar multibarrier structure in graphene. (a) The graphene sheet is deposited on alternating substrates, such as SiO$_{2}$ (red to blue slabs) and SiC (orange slabs), corresponding to wells and barriers, respectively. The red and blue indicate the hot and cold side of the thermoelectric structure. (b) The potential profile of the structure, the blue-red and orange indicate the well and barrier regions, respectively. $k$ and $q_{j}$ are the wave vectors in the wells and barriers, respectively. \end{normalsize}}
    \label{fig:mesh1.}
\end{figure}
\begin{figure}[!t]
    \centering
    \includegraphics[width=7cm, height=6cm]{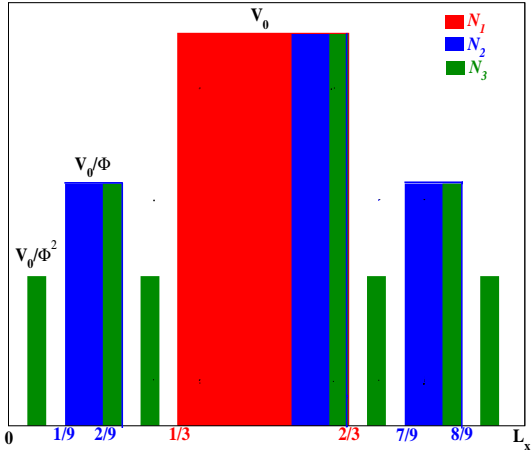}
    \caption{Self-similar barriers of generations ($N_{1}$), ($N_{2}$), and ($N_{3}$). The first generations are shown in red, blue and green, respectively. The red represents the first generation ($N_{1}$) with one barrier, the blue represents the second generation ($N_{2}$) with three barriers and the green $N_{3}$ with seven barriers.    }
    \label{fig:mesh1}
\end{figure}

\begin{equation}
 [ v_{F} ( \vec { \sigma }. \vec {p} )+ V_{0} \sigma_{z}]  \psi   = E \psi,
 \label{eq:1}
\end{equation}
where $\sigma_{i}$ with $i=x,y,z$ is the $ith$ Pauli matrix, $v_{F}$ is the Fermi velocity and $V_{0}=m v^{2}_{F} $ is the mass term, and $\overrightarrow{p}$ is the momentum.
\\ By solving (Eq. (\ref{eq:1})) we can obtain the corresponding eigenfunctions and eigenvalues. In this case, $\psi^{\pm}_{q_{j}}(x,y)$ represents the wave function in the barrier regions, namely:
\begin{equation}
  \psi^{\pm}_{q_{j}}(x,y)=\dfrac{1}{\sqrt{2}} \begin{pmatrix}
1 \\
v^{\pm}_{j}
\end{pmatrix} e^{\pm i q_{x,j}x + i q_{y,j} y}, 
 \end{equation}
where $q_{j}$ is the wave vector, with $j=1,2,..$ and $v^{\pm}_{j}$ are the wave function components given by
 \begin{equation}
  v^{\pm}_{j}=\dfrac{E-V_{0,j}}{\hbar v_{F} (\pm q_{x,j} - i q_{y,j})}, 
  \end{equation}
with the corresponding dispersion relation:
  \begin{equation}
  E^{2}=\pm \hbar^{2} v^{2}_{F}(q^{2}_{x,j}+q^{2}_{y,j})+V_{0,j}.
  \end{equation}
In the case of the well regions, the wave function is given as 
\begin{equation}
  \psi^{\pm}_{k}(x,y)=\dfrac{1}{\sqrt{2}} \begin{pmatrix}
1 \\
u^{\pm}
\end{pmatrix} e^{\pm i k_{x}x + i k_{y} y}, 
 \end{equation}
 where $k$ is the wave vector and  $u^{\pm}$ the corresponding wave function components given by:
\begin{equation}
 u^{\pm}=\dfrac{E}{\hbar v_{F} (\pm k_{x} - ik_{y})},
  \end{equation}
with dispersion relation:
\begin{equation}
E= \pm \hbar v_{F} k.
\end{equation}
To calculate the transmitted wave amplitude $A_{trans}$ and the incident wave amplitude $A_{incid}$, we use the continuity condition of the wave function in each well-barrier interface as well as the conservation of momentum along $y$-coordinate  
$q_{y,j}=k_{y}$. In specific, we can relate the mentioned amplitudes through the so-called transfer matrix method \cite{markos2008wave} :
\begin{equation}
\begin{pmatrix}
A_{incid} \\
B_{incid}
\end{pmatrix} 
=M 
\begin{pmatrix}
A_{trans} \\
0
\end{pmatrix},
\end{equation}
with 
\begin{equation}
 M(E,\theta) =  \begin{pmatrix}
M_{11} & M_{12} \\
M_{21} & M_{22}
\end{pmatrix} 
\end{equation}
the transfer matrix of the system. Once the transfer matrix is known the transmission coefficient can be computed straightforwardly as:
\begin{equation}
 T(E, \theta)= \left| \dfrac{A_{trans}}{A_{incid}} \right|^{2}=\dfrac{1}{\left|  M_{11} \right| ^{2}} , 
\end{equation}
where $M_{11}$ is the first element of the transfer matrix.
The transmission coefficient allows us to calculate the conductance directly through the Landauer-Büttiker formalism \cite{datta1997electronic}. Within this formalism, the conductance for monolayer graphene can be written as:
\begin{equation}
\label{eq:Cond}
 \mathbb{G}(E^{*}_{F})= \dfrac{G}{ G_{0}}= E^{*}_{F} \int_{- \frac{ \pi}{2}}^{ \frac{ \pi}{2}} T(E^{*}_{F}, \theta) \cos \theta d \theta ,  
 \end{equation}
where $ E^{*}_{F} = \dfrac{E_{F}}{V_{0}} $ is the Fermi energy normalized to the height of the main barrier ($V_{0} $), $ G_{0}=\dfrac{2 e^{2} L_{y} V_{0}}{ \hbar^{2} v_{F}  } $ is the conductance fundamental factor, $L_{y}$ is the width of the system in the transverse direction $y$ and $ \theta $ is the angle of incidence of electrons with respect to the propagation direction $x$.
Finally, the thermoelectric properties, particularly the Seebeck coefficient is obtained with the well-known Cutler-Mott formula \cite{cutler1969observation}:
\begin{equation}
 S(E^{*})=S_{0} \dfrac{\partial[\ln G(E^{*})]  }{\partial E^{*}} \bigg|_{ E^{*}=E^{*}_{F}} ,
\end{equation}
with $S_{0}=  \dfrac{\pi^{2} k^{2}_{B} T}{3eV_{0}}$, being $k_{B}$ the Boltzmann constant, $e$ the bare electron charge and $T=50$K is the average temperature between the hot and cold side. The choice of temperature was made in such a way, first, two assure the validity of the Cutler-Mott and second to be in a reasonable range of quantum coherent transport \cite{lunde2005mott}. 
\\Knowing the conductance and the Seebeck coefficient we can compute easily the so-called power factor by multiplying directly the mentioned quantities: 
\begin{equation}
S^{2} G.
\end{equation}
\section{Results and discussion}
The aim of this work is to find out the self-similarity properties, and hence, the scaling rules between generations, in the transmission probability, conductance, Seebeck coefficient and power factor of a new self-similar multibarrier system based on monolayer graphene. Also, we present an important analytical demonstration for the self-similar patterns of the Seebeck coefficient.
In the first place, we start to investigate the transmission scaling rule as a function of the angle for a fixed energy. 
\begin{figure*}[!ht]
\centering
\subfigure{%
\includegraphics[width=8cm, height=7cm]{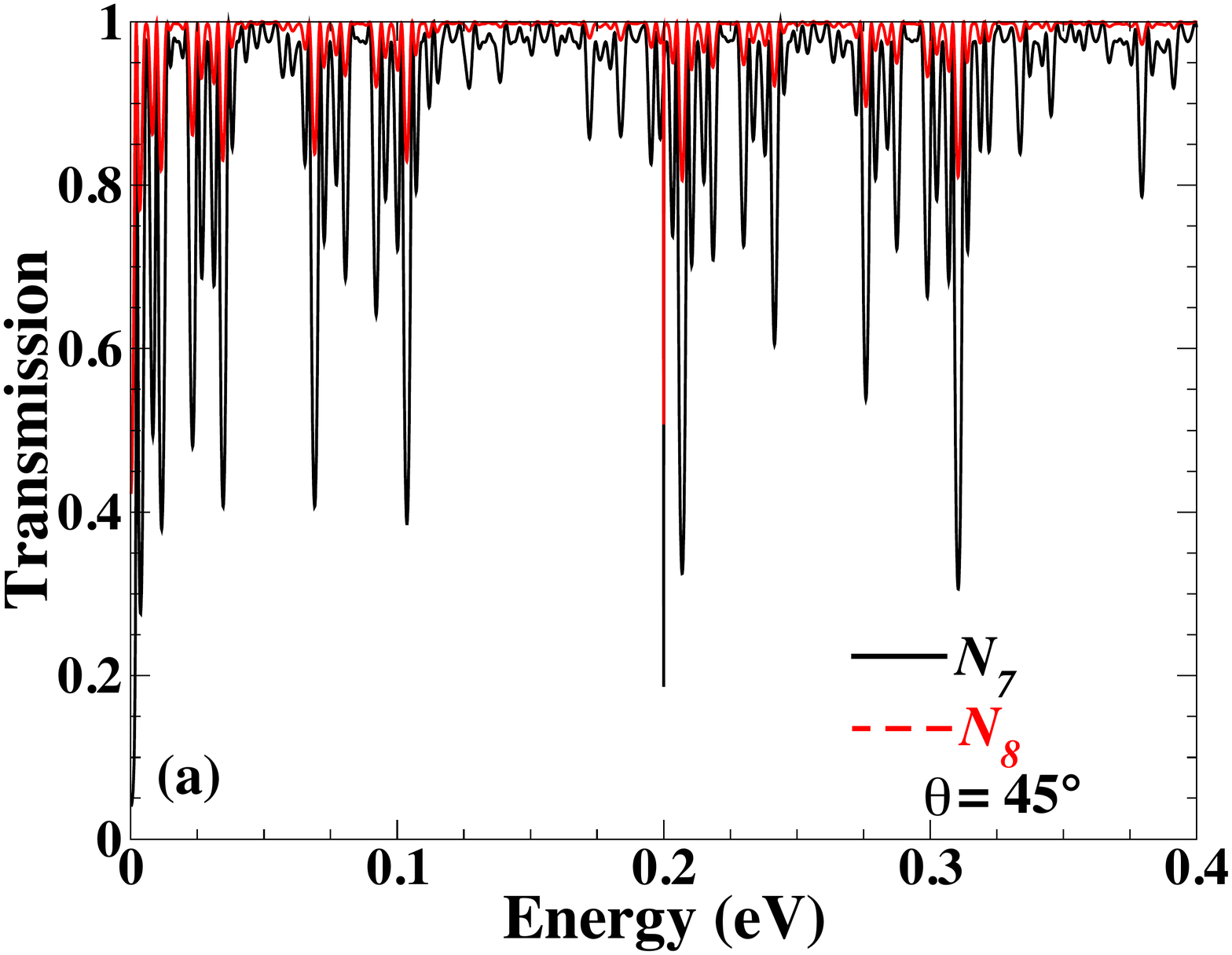}}%
\subfigure{%
\includegraphics[width=8cm, height=7cm]{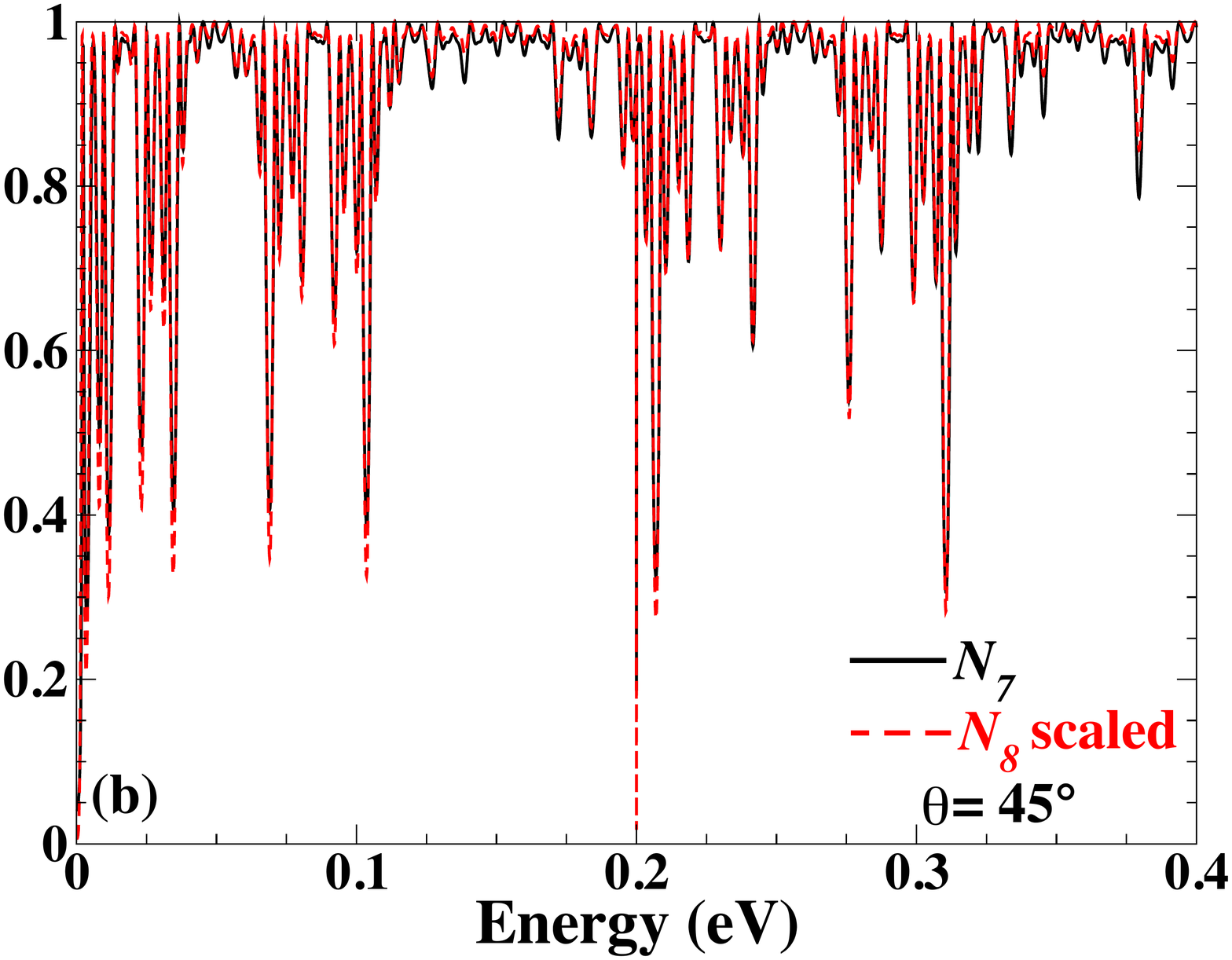}} \\
\hspace{4.00mm}%
\vspace{-4.00mm}%
\subfigure{%
\includegraphics[width=8cm, height=7cm]{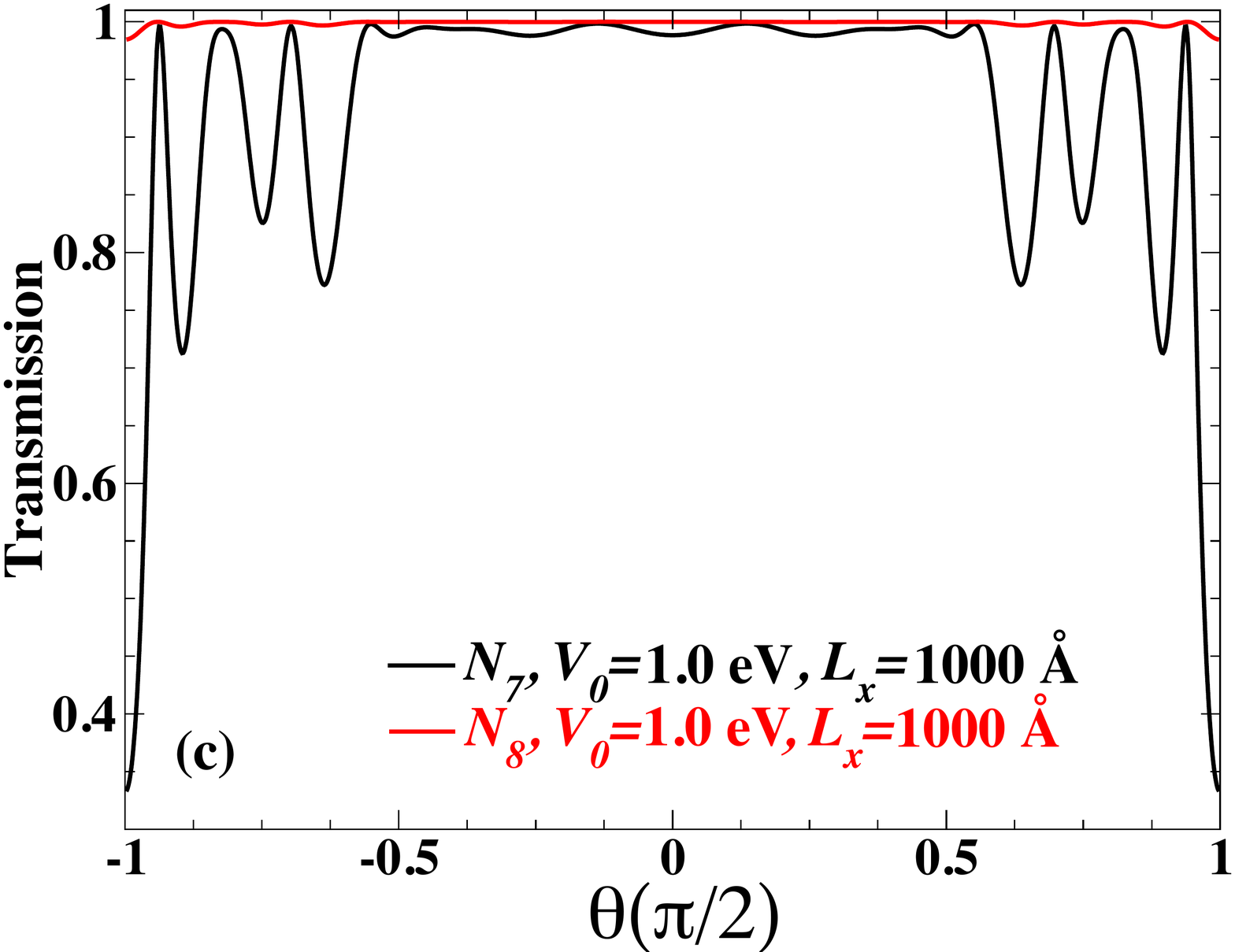}}
\hspace{-0.2cm}%
\vspace{-4.00mm}%
\subfigure{%
\includegraphics[width=8cm, height=7cm]{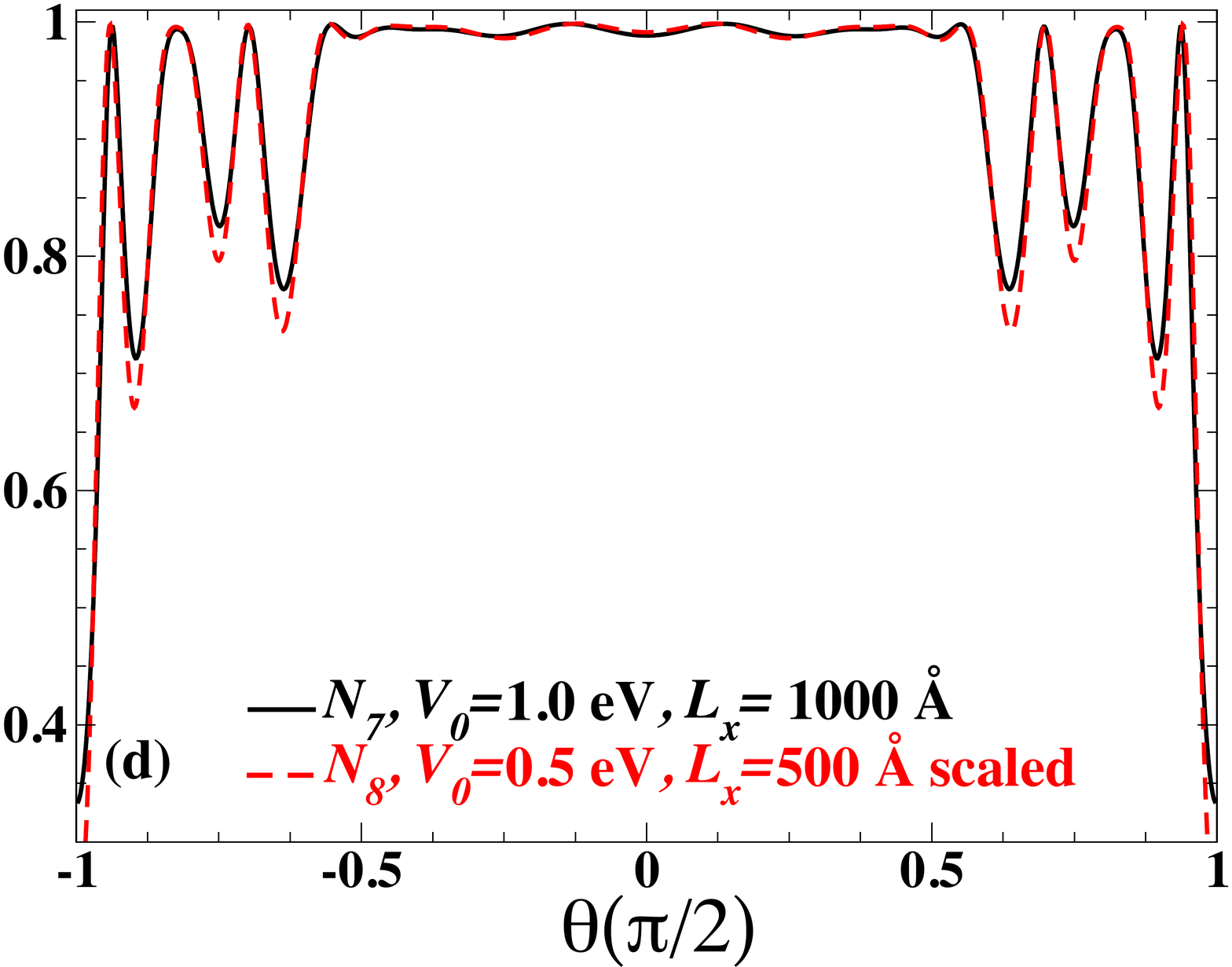}}
\vspace{4.00mm}%
\caption[]
{ Transmission scaling as a function of energy and as a function of the angle of incidence. (a) Transmission patterns as a function of energy at a fixed angle of the incident, $\theta=45^{\circ}$. (b) is the same as (a), but in this case the red curve corresponding to $N_{8}$ is scaled. (c) Transmission patterns as a function of incident angle at a fixed electron energy $E_{x}=0.15$ eV. (d) the same as (c) but in this case, the red curve corresponding to $N_{8}$ is scaled. In (a) and (b), the height of main barrier is $V_{0}=0.2$ eV and the length of system is $L_{x}=10000$ \AA. }%
\end{figure*}
\begin{figure*}[!ht]
\centering
\subfigure{%
\includegraphics[width=8cm, height=7cm]{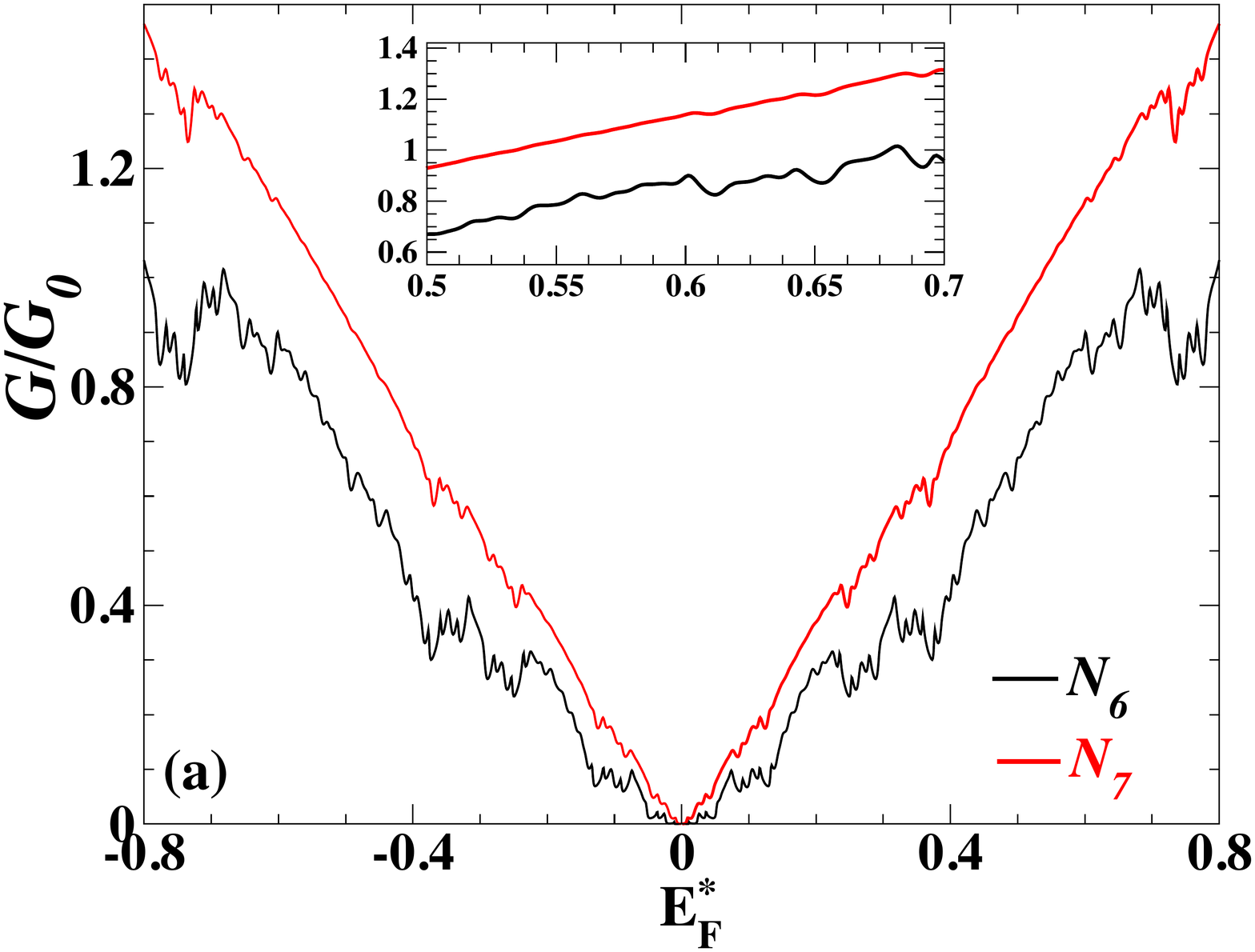}
}%
\hspace{4.00mm}%
\vspace{-4.00mm}%
\subfigure{%
\includegraphics[width=8cm, height=7cm]{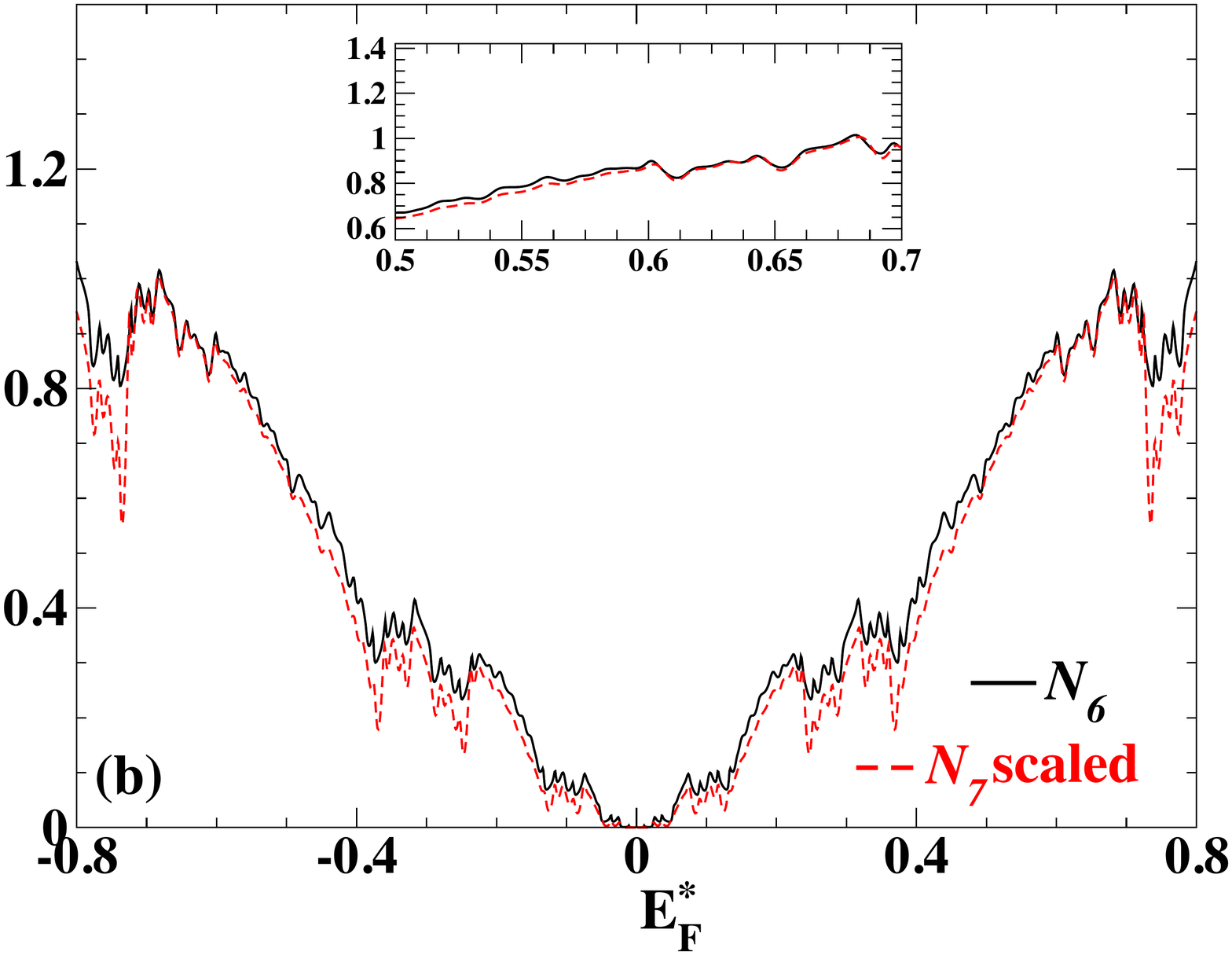}
} \\
\subfigure{%
\includegraphics[width=8cm, height=7cm]{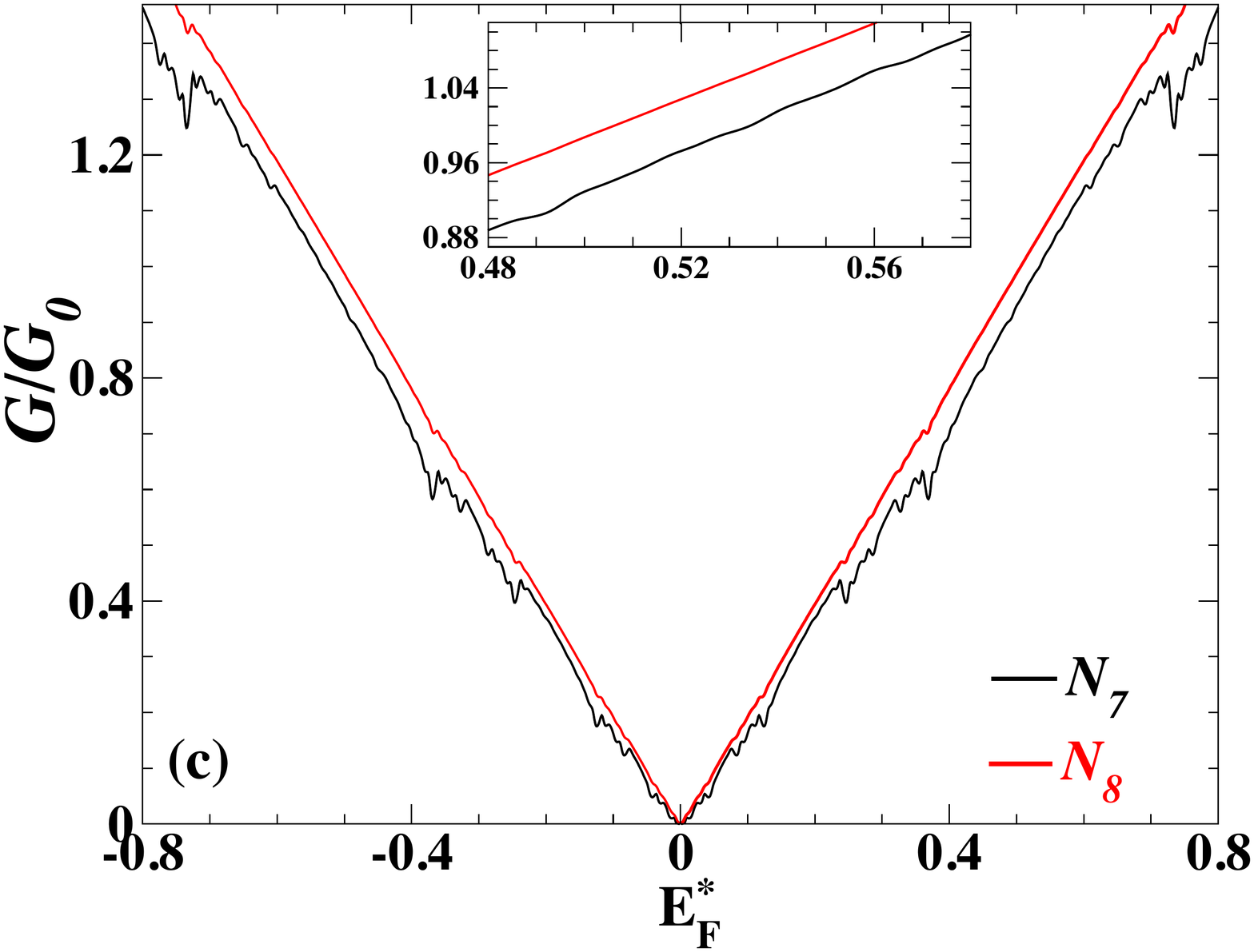}
}%
\hspace{4.00mm}%
\vspace{-4.00mm}%
\subfigure{%
\includegraphics[width=8cm, height=7cm]{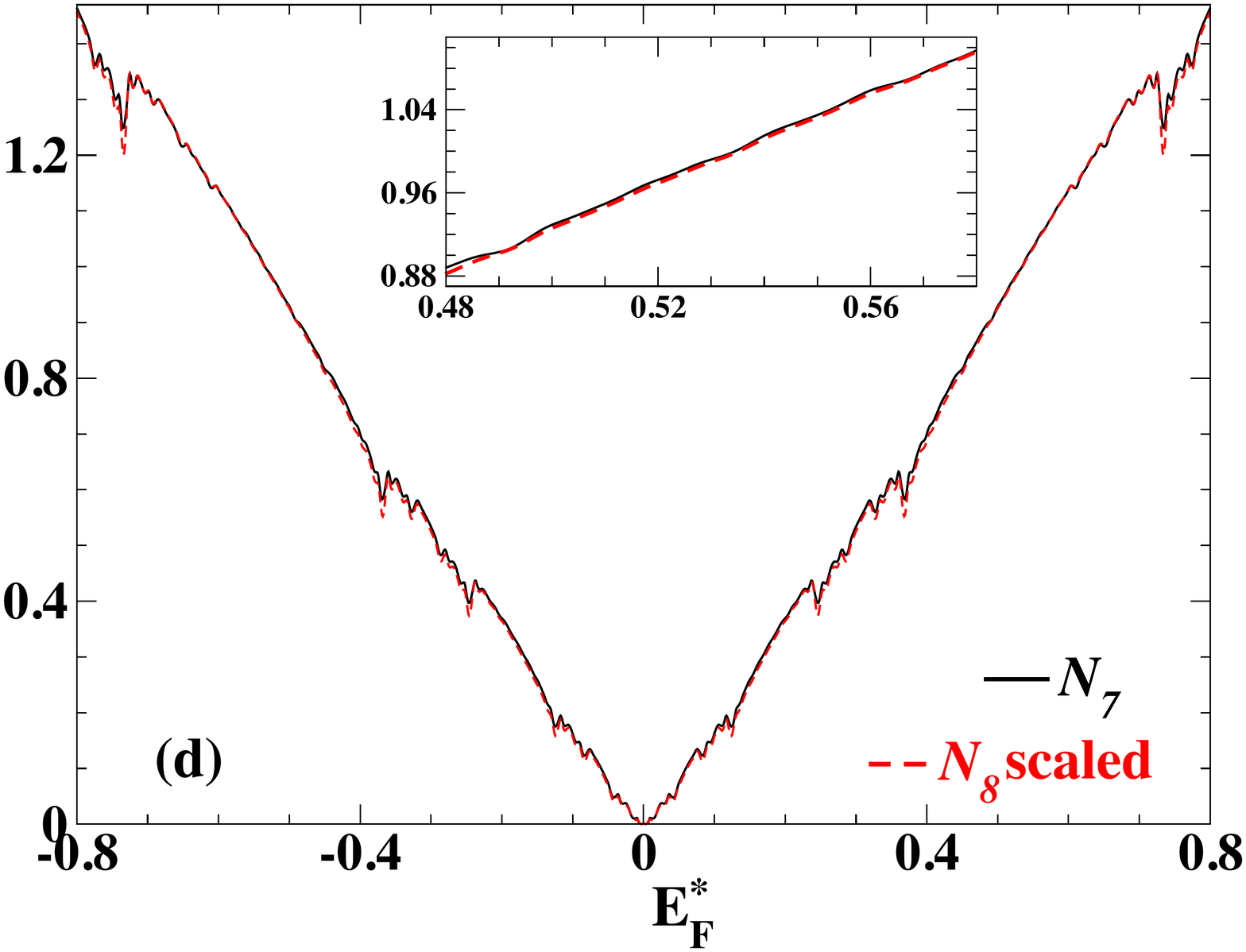}
}%
\caption[]
{Scaling conductance $G$ between the generations $N_{6}$, $N_{7}$, $N_{8}$ as a function of the Fermi energy. (a), (c) Respectively, conductance curves between pairs of generations ($N_{6}, N_{7}$) and ($N_{7}, N_{8}$). (b), (d) Same curves but in this case, generation $N_{7}$ (red curve) and $N_{8}$ (red curve) are scaled. In the inset figure we appreciate the overlapping of patterns in a reduced interval of Fermi energy. The height of main barrier $V_{0}=0.2$ eV, and the total length of the system is $L_{x}=10000$ \AA. }%
\end{figure*}
In Figs. 3(a)-3(b), we present scaling properties of transmission at a fixed incidence angle, $\theta=45^{\circ}$. Fig. 3(a), displays the transmission as a function of energy for generations $N_{7}$, $N_{8}$ with a specific height of the main barrier $V_{0}=0.2$ eV and total length of the system $L_{x}=10000$ \AA. Fig. 3(b), presents the scaled transmission $N_{8}$, with parameters different from those of the curve $N_{7}$, as we can see, both curves, $N_{7}$ and $N_{8}$ are overlapped, this overlapping occurs when we apply a mathematical transformation. The transformation comes as:
 $T_{N_{7}}(\theta,E_{x}) \approx [T_{N_{8}} (\theta,E_{x})]^{6}$, we name this transformation the scaling rule. Expressed otherwise, the transmission curves are related by a certain power, but differ only in their corresponding generation. The scaling rule is a mark of self-similarity in the transmission properties as in transmission at normal incidence \cite{diaz2015scaling,diaz2016self,rodriguez2016self}. In this kind of system, self-similar properties practically arise from the generation $N_{7}$, because the system begins to exhibit a nearly self-similar behavior from $N_{6}$. This scaling rule is valid for each generation pairs, which leads us to generalize it:
\begin{equation}
T_{N}(\theta,E_{x}) \approx [T_{N+p} (\theta,E_{x})]^{\eta^{p}},
\end{equation}
where $N$ is the generation number, $\eta$ is the scale factor whose value is equal $6$, and $p$ is the difference between generations.
In Figs. 3(c)-3(d), we present scaling properties of transmission as a function of the angle of incidence. Fig. 3(c) presents the transmission as a function of incident angle for generations $N_{7}$, $N_{8}$ with a specific height of the main barrier $V_{0}=1.0$ eV and total length of the system $L_{x}=1000$ \AA. Fig. 3(d), presents the transmission properties with mentioned parameters, however the generation $N_{8} $ is scaled by $V_{0}=0.5$ eV and $L_{x}=500$ \AA, as well as a mathematical transformation : $T_{(7;1.0;1000)}(\theta,E_{x}) \approx [   T_{(8;0.5;500)} (\theta,E_{x})]^{5}$. According to Fig . 3(d) it is clear that the transmission of both generations $N_{7}$ and $N_{8} $ scaled, are overlapped,  these scaling properties can be expressed as a scaling rule in a generalized way: 
\begin{equation}
 T_{   (N;V_{0};L_{x})}(\theta,E_{x})  \approx    [   T_{( N+p; V_{0} / \beta
; L_{x} / \gamma)} (\theta,E_{x})]^{\eta^{p}(\gamma\beta)^{2}},
\end{equation}
where $N$ is the generation number, $\eta$ is the scale factor whose value is equal $5$, $p$ is the difference between generations, and the factors $\gamma$ and $\beta$ come in multiples of two.
\\In fact, this analytical expression is an approximation (combination rule) of the individual scaling rules between generations, the energy of the main barrier, and the length of the total system.  
In order to compare the matching between non-scaled and scaled curves, we calculate the root-mean-square-deviation (RMSD) for Figs. 3(b)-3(d). The results are presented in table 1.
\begin{table*}[!ht]
\caption{\begin{normalsize}
The root-mean-square-deviation of the transmission scaling.
\end{normalsize}} 
\centering 
\begin{tabular}{c rrrrrrr} 
\hline\hline 
\multicolumn{7}{c}{RMSD} \\ [0.5ex]
\hline 
Scale factor $\eta$ & $4$ & $5$ & $6$ & $7$& $8$& $9$\\
Figure. 3(b) & $3.98x10^{-2}$  & $2.79x10^{-2}$  & $2.18x10^{-2}$ & $2.39x10^{-2}$ & $3.15x10^{-2}$ & $4.10x10^{-2}$ \\ 
Figure. 3(d) & $3.13x10^{-2}$  & $1.75x10^{-2}$ & $2.15x10^{-2}$ & $3.51x10^{-2}$ & $4.97x10^{-2}$ & $6.39x10^{-2}$ \\[1ex] 
\hline 
\end{tabular}
\end{table*}
\\ The scale factor $\eta$, with lower RMSD gives the best approximation between the scaled and non-scaled curve. Hence, it is clear that the best scale factor value $\eta$ is 6 for scaling rule (Eq. (14)). The best scale factor $\eta$ is $5$ for scaling rule (Eq. (15)).  
\par In the second place, the transmission allows to calculate directly the conductance by summing over all transmission channels, see (Eq. (\ref{eq:Cond})). So, it is possible to explore the self-similarity and the scalability of the conductance patterns. In Fig. 4, we show the conductance patterns for generations $N_{6},N_{7}$ and $N_{8}$ as a function of Fermi energy. In particular, we have considered pairs of generations ($N_{6},N_{7}$) and ($N_{7},N_{8}$), (Fig. 4(a)-4(c)), respectively. At first sight, we notice that the two curves diverge when the Fermi energy increases. However, with an appropriate mathematical transformation, we can connect the pairs of conductance curves ($N_{6},N_{7}$) and ($N_{7},N_{8}$), (see Figs. 4(b)-4(d)). The mathematical transformation comes as: $\mathbb{G}_{7} (E^{*}_{F} )\approx [\mathbb{G}_{8} (E^{*}_{F})]^{6}/(2E^{*}_{F} )^{5}$ , where the subscript is the generation number. It is interesting to note that the factors $6$ and $(2E^{*}_{F} )^{5}$ are the scale factors that relate the generations. In other words, this transformation shows that there is a self-similar behavior. We can also generalized this transformation as: 
\begin{equation}
\mathbb{G}_{N} (E^{*}_{F} )\approx \dfrac{[\mathbb{G}_{N+p} (E^{*}_{F})]^{\eta^{p}}}{(2E^{*}_{F} )^{\eta^{p}-1}},
\end{equation}
where $N$ is the generation number, $\eta$ is the scale factor whose value is equal $6$ and $p$ is the difference between generations.
\\We have also computed the RMSD for the scaling conductance Fig. 4(d), showing the results in table 2.
\begin{table*}[!ht]
\caption{ \begin{normalsize}
The root-mean-square-deviation of the conductance scaling. \end{normalsize}} 
\centering 
\begin{tabular}{c rrrrrrr} 
\hline\hline 
\multicolumn{7}{c}{RMSD} \\ [0.5ex]
\hline 
Scale factor $\eta$ & $4$ & $5$ & $6$& $7$& $8$& $9$\\
Figure. 4(d)  &$2.95x10^{-2}$ &  $1.28x10^{-2}$  & $8.10x10^{-3}$  & $2.32x10^{-2}$  & $3.94x10^{-2}$ & $5.55x10^{-2}$   \\ 
\hline 
\end{tabular}
\end{table*}
Based on the results of RMSD, we can note that the best scale factor value $\eta$, in the case of scaling conductance rule (Eq. (16)) is $6$. 
Even more interesting, the scaling properties in the case of conductance appear further between different heights of the main barrier and the total lengths of the system (see supplementary information). 
\begin{figure*}[!ht]
\centering
\subfigure{%
\includegraphics[width=8cm, height=7cm]{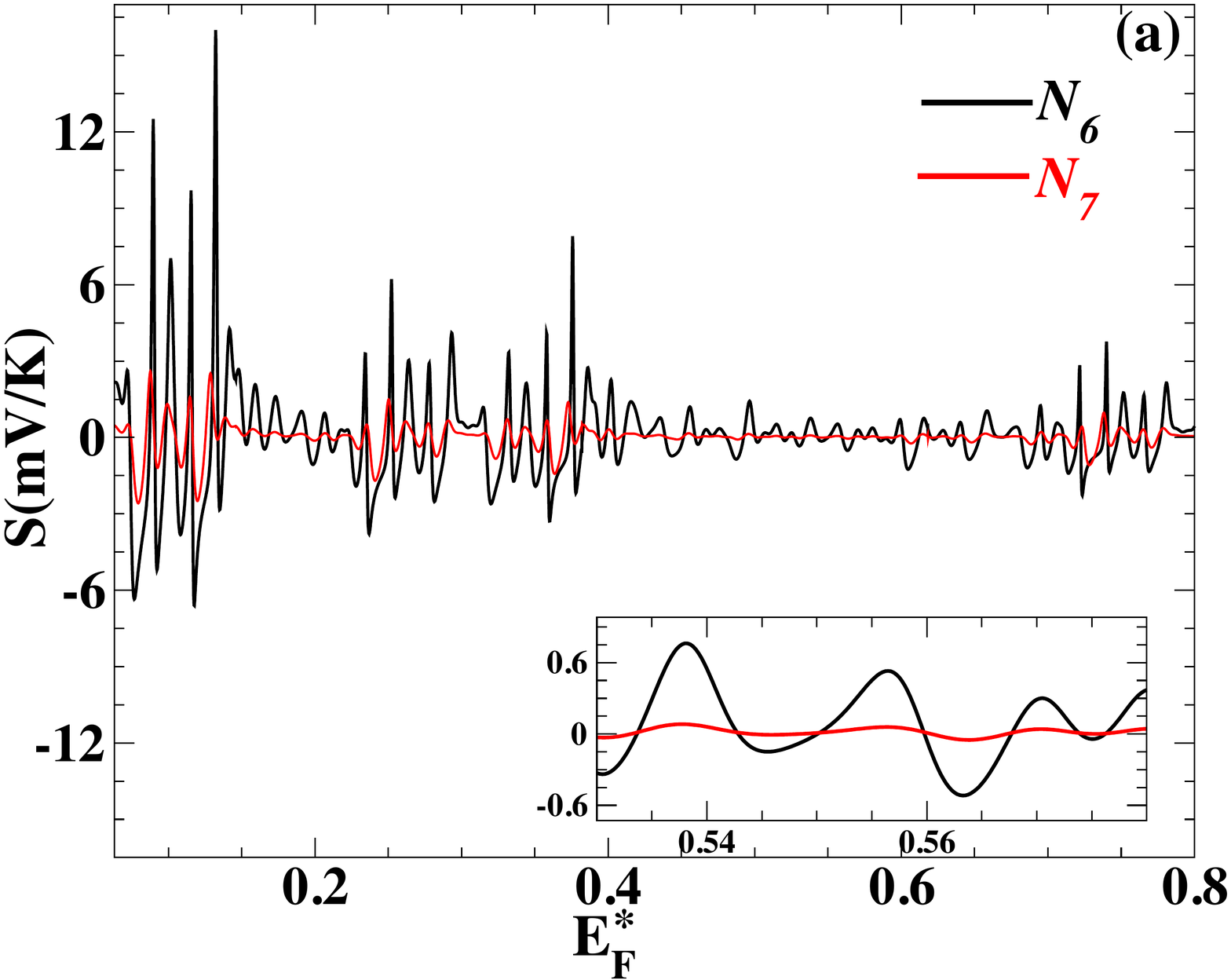}}%
\hspace{4.00mm}%
\vspace{-4.00mm}%
\subfigure{%
\includegraphics[width=8cm, height=7cm]{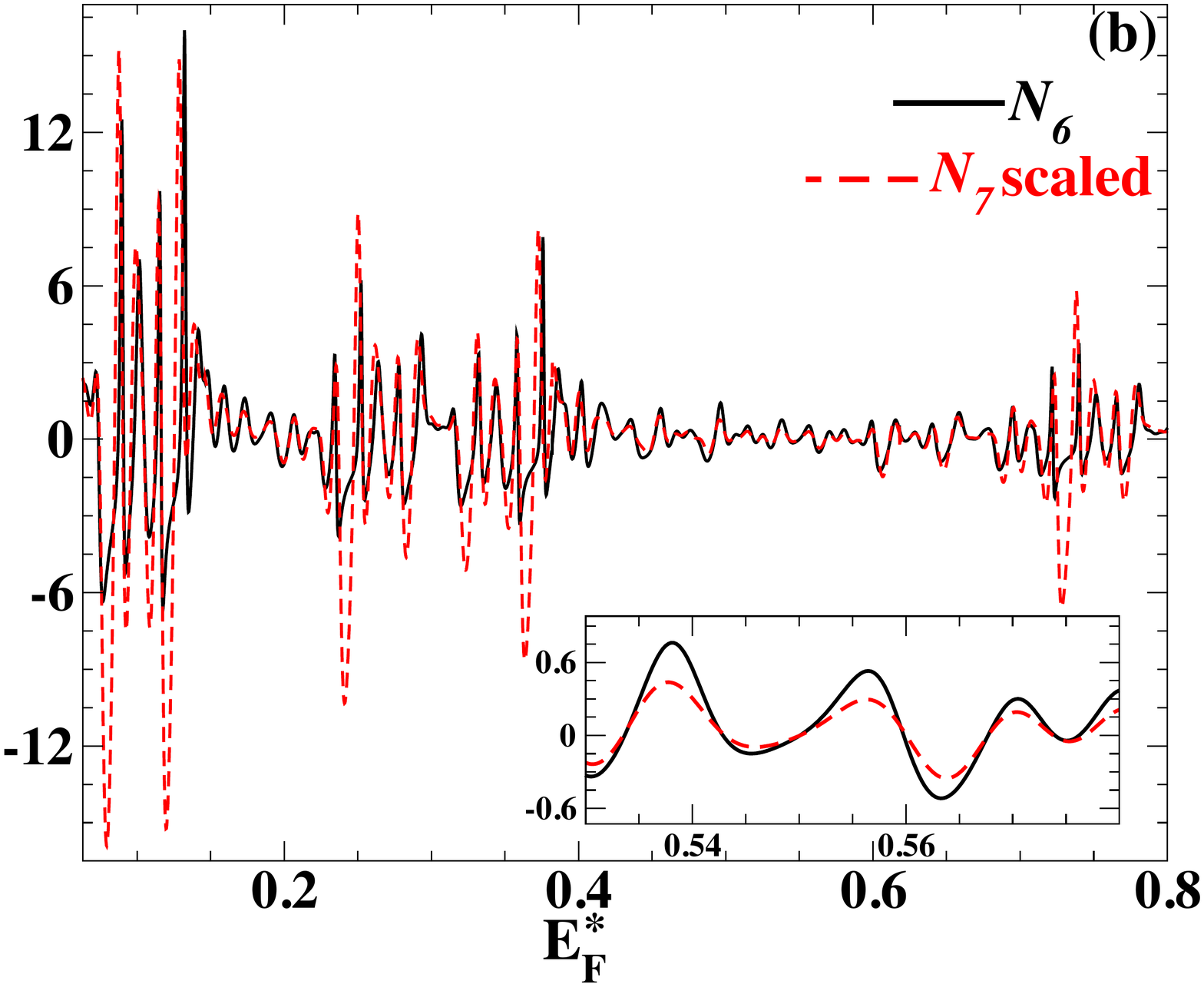}} \\
\subfigure{%
\includegraphics[width=8cm, height=7cm]{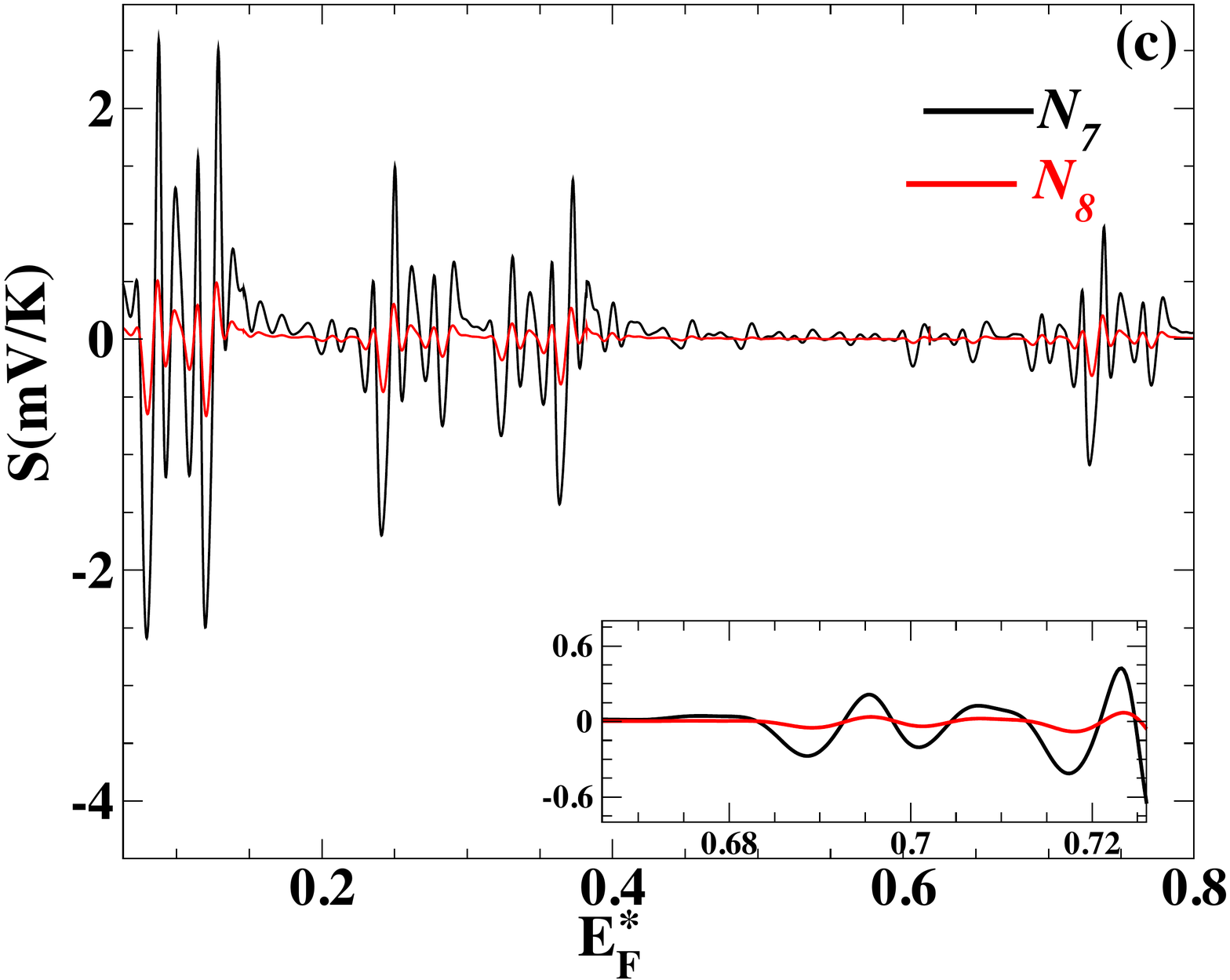}}%
\hspace{4.00mm}%
\vspace{-4.00mm}%
\subfigure{%
\includegraphics[width=8cm, height=7cm]{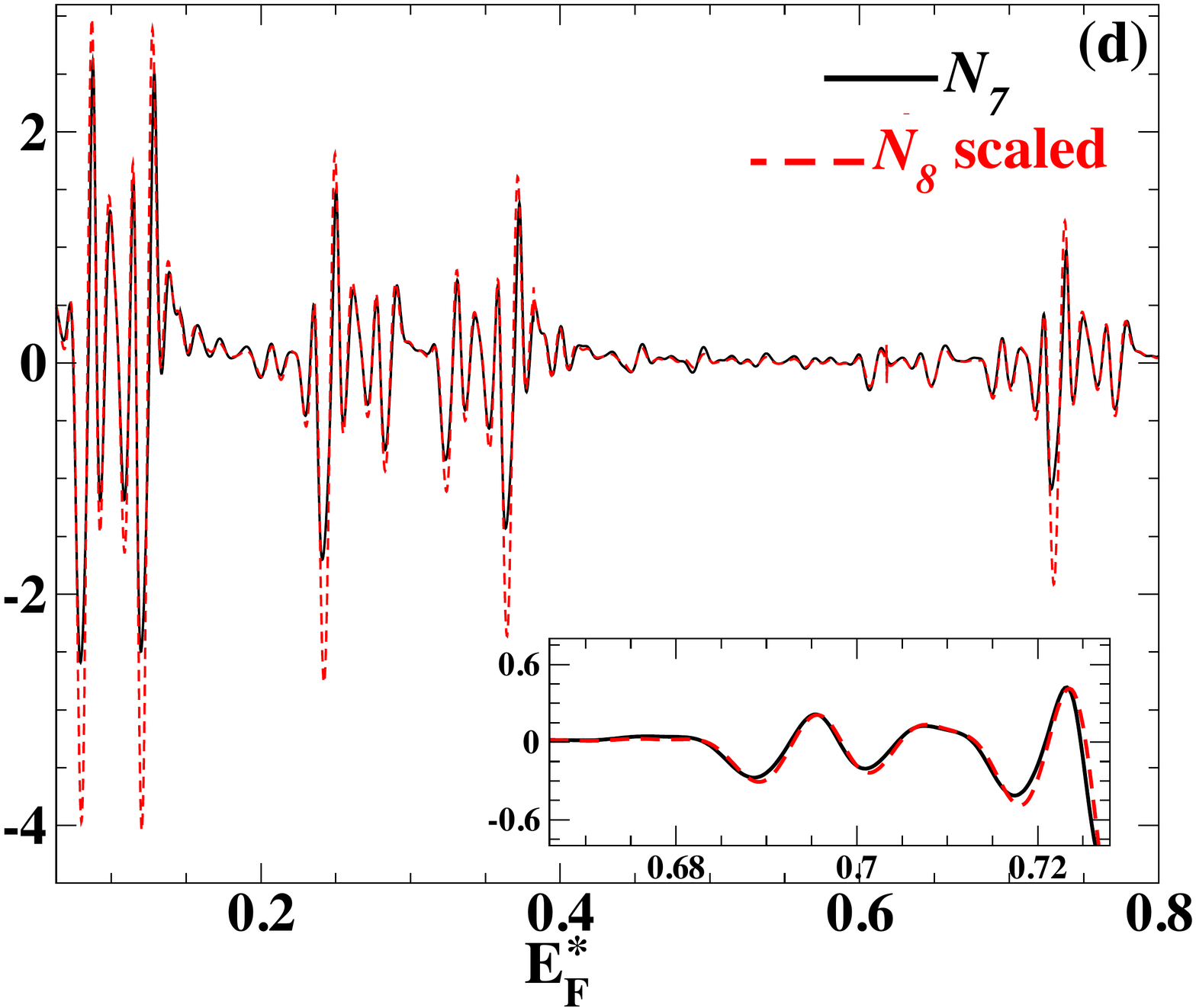}}%
\caption[]
{Seebeck coefficient between generations $N_{6}, N_{7}, N_{8}$ as a function of the Fermi energy. (a), (c) Represent respectively the curves of Seebeck coefficient between the generations ($N_{6}, N_{7}$) and ($N_{7}, N_{8}$). (b), (d) are the same curves but in this case, both pairs of generations are modified by a scaling transformation (see text). In the inset figure we appreciate the overlapping of patterns in a reduced interval of Fermi energy. In this instance, the height of main barrier is $V_{0}=0.2$ eV, and the total length of the system is $L_{x}=10000$ \AA.  }%
\end{figure*}
\\In Figs. 5(a)-5(c), we report the Seebeck coefficient as a function of Fermi energy, for generations $N_{6}, N_{7}$ and $N_{8}$. Moreover, as in the case of the conductance and transmission coefficient, we investigate the Seebeck scaling rule between generations. In Figs. 5(b)-5(d), we expect that the same thing happens between ($N_{6}, N_{7}$) and ($N_{7}, N_{8}$). In particular, we explore the following scaling rule: $S_{6} \approx 6 S_{7}- 5 S_{0}  /E^{*}_{F} 
$, where both $6$ and $5 S_{0}  /E^{*}_{F}$ represent the scale factors. As we can see this scaling rule works quite well. It is important to remark that the scaling rule for the Seebeck coefficient is remarkably different from those found for the transmission and conductance. As in the case of the transmission and conductance, we propose a general expression that connects the Seebeck coefficient patterns between generations: 
 \begin{equation}
S_{N}(E^{*}_{F})\approx \eta^{p}  S_{N+p}(E^{*}_{F}) - \dfrac{(\eta^{p}-1) S_{0}}{E^{*}_{F}}, 
\end{equation}
where $p$ is the difference between generations, $\eta$ is the scale factor whose value is equal $6$, and $N$ is the generation number. 
\\The RMSD of scaling Seebeck coefficient is calculated and shown in table 3, according to RMSD the scale factor $\eta$ with lower RMSD is $3$, so in a Seebeck scaling rule (Eq. (17)) it has to be $3$ not $6$ which presents the best scaling in this case.  
\begin{table*}[!ht]
\caption{\begin{normalsize} The root-mean-square-deviation of the Seebeck coefficient scaling. \end{normalsize}} 
\centering 
\begin{tabular}{c rrrrrrr} 
\hline\hline 
\multicolumn{7}{c}{RMSD} \\ [0.5ex]
\hline 
Scale factor $\eta$ & $3$ & $4$ & $5$ & $6$ & $7$ & $8$ & $9$ \\
Figure. 5(d)   & $1.95x10^{-3}$ & $2.02x10^{-3}$ & $2.32x10^{-3}$  & $2.78x10^{-3}$  & $3.34x10^{-3}$  & $3.95x10^{-3}$ & $4.60x10^{-3}$  \\  
\hline 
\end{tabular}
\end{table*}

\begin{figure*}[!ht]
\centering
\subfigure{%
\includegraphics[width=8cm, height=7cm]{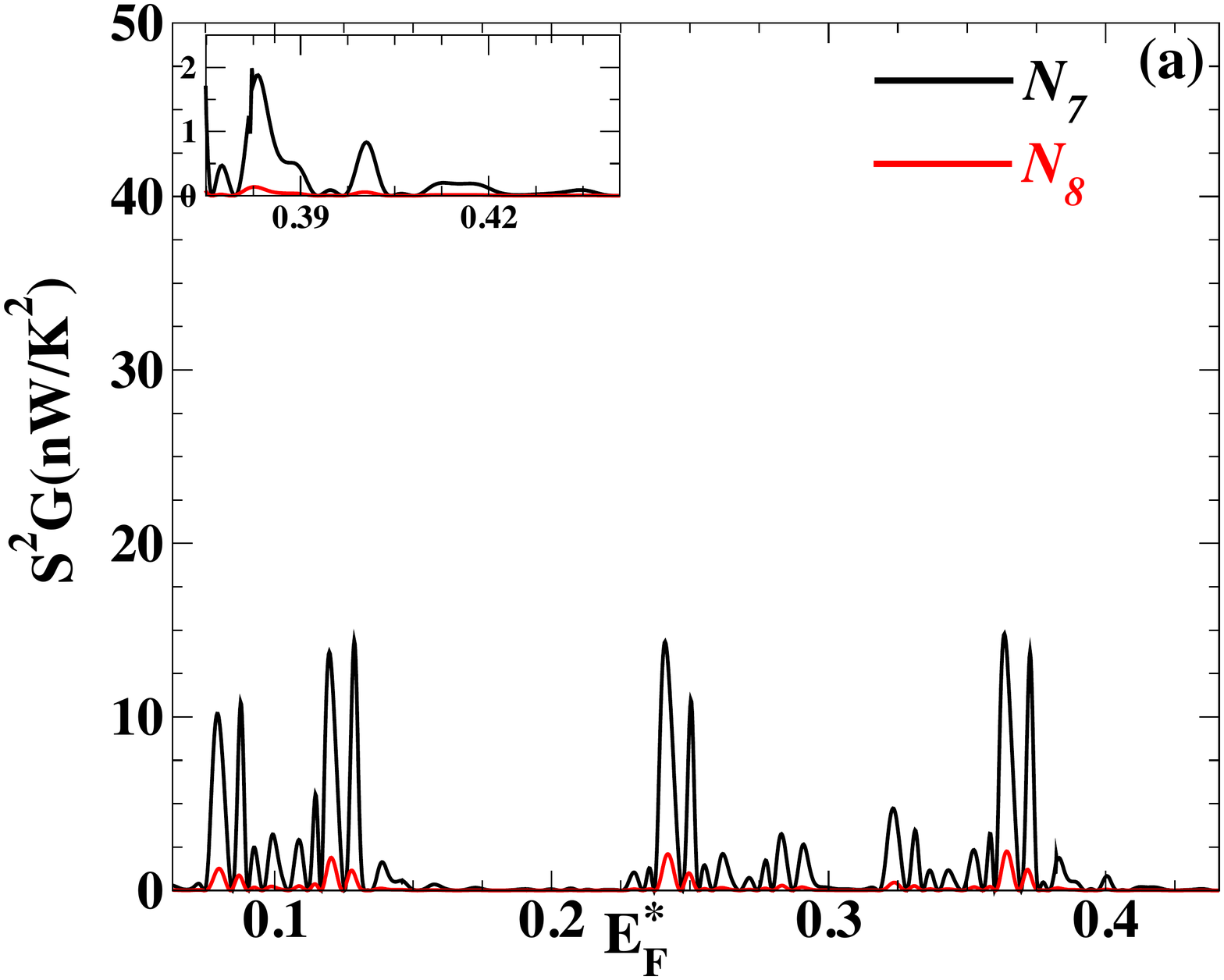}}%
\hspace{0.5mm}%
\subfigure{%
\includegraphics[width=8cm, height=7cm]{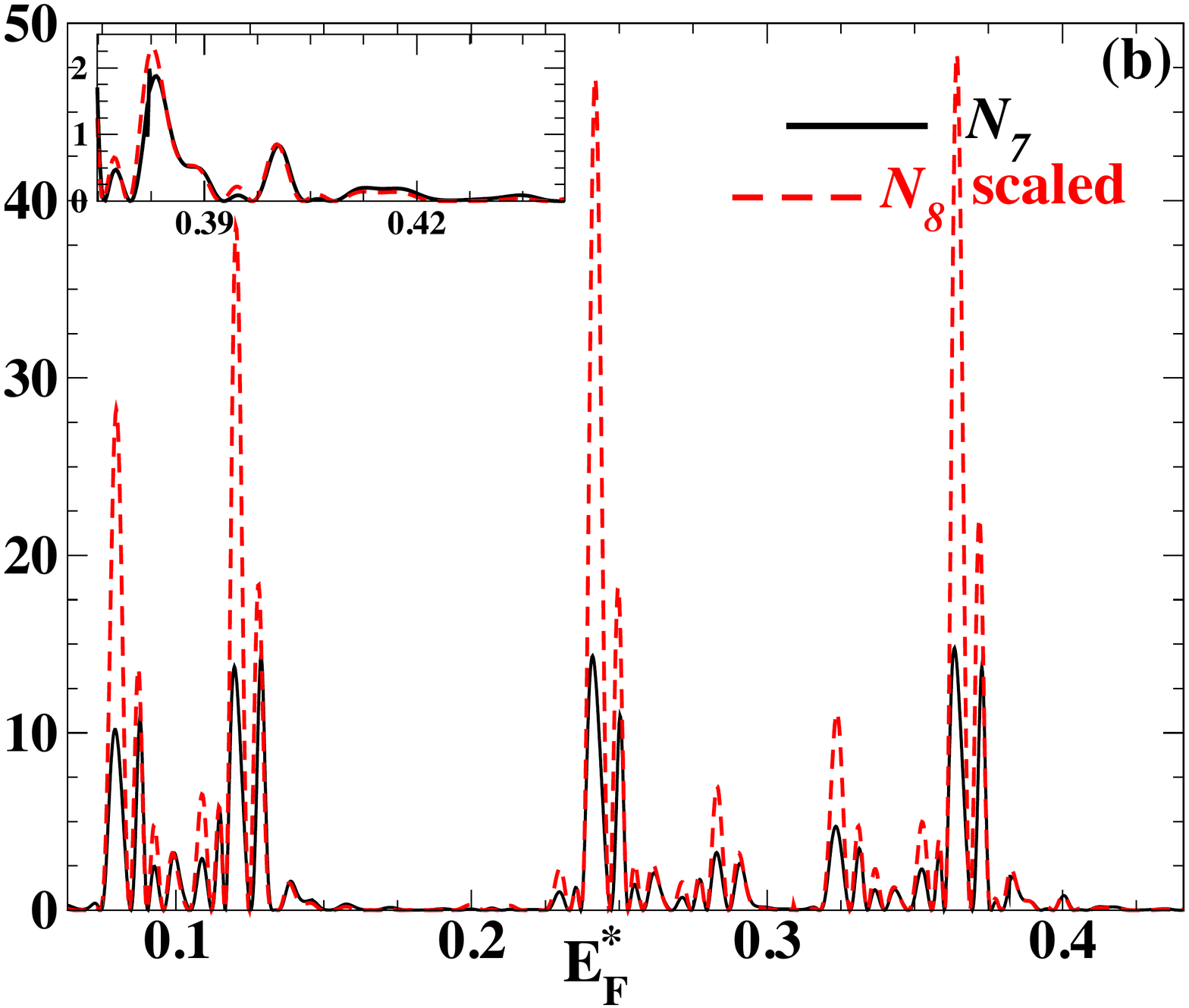}} \\
\vspace{-4.00mm}%
\caption[]{Scaling power factor between generations $N_{7}, N_{8}$ as a function of the Fermi energy. (a) Power factor between generations ($N_{7}, N_{8}$). (b) Same generations but in this case, generation $N_{8}$ is scaled by the scaling transformation (Eq. (23)). In the inset figure we appreciate the overlapping of patterns in a reduced interval of Fermi energy. For instance, the height of main barrier $V_{0}=0.2$ eV, and the total length of the system is $L_{x}=10000$  \AA . }%
\end{figure*}
Now, it is turn to talk about the conditions to find scaling rules for the transmission coefficient, the conductance and the Seebeck coefficient in this kind of system: according to our results and to other investigations, it is important to have a self-similar system as well as Dirac electrons \cite{diaz2016self, rodriguez2016self}.
\\ In the case of conductance and Seebeck coefficient, scaling rules are different compared to that in transmission coefficient because scaling rules in this case cannot be found without scaling the energy, which also makes it very difficult to find these rules. 
Another aspect that we want to discuss is the validity of the scaling rules. In relation to this point, it is interesting to note that our system has limits. For example, at low generations our system is not really self-similar, so we do not expect that scaling rules arise in transmission coefficient and conductance, neither in Seebeck coefficient. At high generations, we have to be careful, because the length of our structure will exceed the carbon-carbon distance, therefore, our system will be out of physical sense. Our system is also limited by the physical design that we propose. For instance, the height of the main barrier cannot be greater than $2$ eV, otherwise the relativistic description in graphene is no longer valid. Regarding the limits that we have mentioned, we can deal with them by adjusting the system parameters, particularly the length of the structure and the height of the main barrier to reach more generations with self-similar characteristics. It is important to have in mind that the scaling rules found in the transmission coefficient cannot be directly implemented in the conductance and  Seebeck coefficient, due to the sum over all transmission channels, which makes not possible to perform an analytical demonstration of the self-similarity patterns. However, we have successfully implemented the conductance directly into the Seebeck coefficient. 
\par Now, we present here the analytical demonstration of self-similar properties in the Seebeck coefficient. From the analytical general expression of conductance (Eq. (16)), we try to find the Seebeck scaling rule, then we compare it with the numerical outcomes.  
Firstly, we write the general formula for Seebeck coefficient:  
\begin{equation}
 S= S_{0}  \dfrac{\partial \ln(\mathbb{G}(E^{*}) )}{\partial E^{*}} \bigg|_{ E^{*}=E^{*}_{F}}. 
 \end{equation}
Then we formulate this equation for generation $N_{7}$ : 
 \begin{equation}
S_{7}= S_{0}  \dfrac{\partial \ln(\mathbb{G}_{7}(E^{*}) )}{\partial E^{*}} \bigg|_{ E^{*}=E^{*}_{F}}. 
\end{equation}
By replacing $N_{7}$ in terms of $N_{8}$ according to the conductance scaling rule (Eq. (16)):
 \begin{equation}
 S_{7}= S_{0}  \dfrac{\partial \ln \left(\dfrac{(\mathbb{G}_{8})^{\eta}}{(2E^{*})^{\eta - 1}}\right)}{\partial E^{*}} \bigg|_{ E^{*}=E^{*}_{F}}.
 \end{equation}
 Deriving this equation yields:
 \begin{equation}
 S_{7}=\eta S_{0}   \dfrac{\partial \ln (\mathbb{G}_{8} )}{\partial E^{*}}  - \dfrac{(\eta-1) S_{0} }{E^{*}} \bigg|_{ E^{*}=E^{*}_{F}},
 \end{equation}
or equivalently:
 \begin{equation}
S_{7}=\eta  S_{8} - \dfrac{(\eta-1) S_{0}}{E^{*}} \bigg|_{ E^{*}=E^{*}_{F}}. 
\end{equation}
\\ As we can notice the numerical and analytical scaling rules are practically the same, compare (Eqs. (17) and (22)). However, the scale factor can be adjusted to give a better scaling. For example, for the case that we have presented, a scale factor 3 can be used instead of 6, giving a better scaling as can be seen directly in the curves of Fig. 5, as well as in the RMSD for the mentioned $\eta$.
\\After analytically and numerically finding the scaling rule for the Seebeck coefficient, we carried out a search for the scaling rules of the power factor using the same method. As we have stated in the methodology, the power factor is the product of the square of the Seebeck coefficient and the conductance $S^{2}G$. In Fig. 6(a), we show the power factor as a function of Fermi energy for generations ($N_{7}, N_{8}$). As we have stated throughout the manuscript the derivation of the scaling rules is in general tricky. So, the power factor is not the exception. In fact, it is really challenging, even numerically. Guided by the Seebeck coefficient results, we will try to find the scaling rule for the power factor by using directly the scaling rules for the conductance and Seebeck coefficient, namely:
\begin{equation}
S_{7}^{2} G_{7}(E_{F}^{*}) \approx \left(6 S_{8}(E_{F}^{*})-\dfrac{5 S_{0} }{E_{F}^{*}}\right)^{2} \dfrac{ [G_{8}(E^{*}_{F})]^{6} G_{0}}{(2E^{*}_{F}  )^{5} }.
\end{equation}
In Fig. 6(b), we show the scaling results for the power factor. As we can notice the scaling works reasonably well despite the complexity of (Eq. (22)).
This gives us the margin to think in a general expression:
\begin{equation}
S_{N}^{2} G_{N}(E_{F}^{*}) \approx  \left(\eta^{p} S_{N+p}(E_{F}^{*})-\dfrac{(\eta^{p} - 1) S_{0} }{E_{F}^{*}}\right)^{2} \dfrac{[ G_{N+p}(E_{F}^{*})]^{\eta^{p}} G_{0}}{(2E_{F}^{*})^{\eta^{p}-1}}.
\end{equation}
\\ Unfortunately, in this case, we cannot relate explicitly the power factors between to specific generations. In relation to this point, we want to remark that the task of finding analytically the scaling rules in the transmission coefficient and the conductance is really difficult. In fact, the transmission coefficient is the result of the transfer matrix, which is the multiplication of several matrices. Likewise, the conductance is the integral of the transmission coefficient in all transmission channels. This is why it is complicated to find analytical expression for the self-similar patterns. Surprisingly, it is possible to do it for the Seebeck coefficient by replacing the scaling expression of the conductance directly into the well-known Cutler-Mott formula. In fact, it works well and agrees with the scaling rule that is numerically obtained. Finally, in regard to the transmission coefficient, conductance and even the power factor, the search of self-similarity rules remains a challenge and it is a perspective of this work.  
\\It is also worth mentioning that in this type of structure the conductance increases as the generation increases (see Fig. 4). This is a general trend in self-similar structures. Actually, the mentioned trend is directly related to the fragmentation of the structure, specifically, the width and height of the barriers. This fragmentation of the barriers favors propagating states, giving place to a general enhancement in the conductance and the Seebeck coefficient \cite{garcia2017self}.  
\\Finally, it is important to remark that the self-similar system that we present in this work could be a challenge for experimentalists. The main obstacle it is that we required a substrate that interacts differently with the graphene sheet \cite{rodriguez2016self, garcia2017self}. Fortunately, other external effects, electric and magnetic fields, that are a reality from the experimental standpoint in 2D materials can be an option to generate self-similar structures.
\section{Conclusions}
To our knowledge, this is the first study exploring thermoelectric effects in a self-similar system based on graphene. We have studied the self-similarity patterns and scalability in the transmission coefficient as a function of the incident angle, conductance, the Seebeck coefficient, and power factor in a new self-similar multibarrier structure, generated by substrates and based on a graphene monolayer. The self-similar barriers in this system are scaled by the golden ratio in their height at each generation. The golden number in this case is one of the few scaling factors that give scaling rules in this structure. In particular, we have implemented scaling rules between generations using general mathematical expressions. Moreover, we present an analytical demonstration of self-similarity in the Seebeck coefficient. These scaling rules represent the self-similar patterns in a thermoelectric system based on graphene. The mentioned system is one of the few presenting self-similar patterns in the physical properties of monolayer graphene.  

\section*{Acknowledgements}
M.M. acknowledges CONACYT-Mexico for the scholarship for doctoral studies. O.O. would like to acknowledge SEP-SES, No. 511-6/2019.-109601 for supporting this work. I.R.-V. is thankful to CONACYT-SEP Mexico for the financial support through grant A1-S-11655.


\bibliographystyle{elsarticle-num}
\bibliography{riaibib.bib}
\medskip

\end{document}